\documentclass[usenatbib,usegraphicx,twocolumn,a4paper]{mn2e}
\usepackage[latin1]{inputenc}
\usepackage{graphicx,amsmath}
\usepackage{color}
\usepackage{times}
\usepackage{amssymb}
\usepackage{natbib}
\usepackage{rotating}
\usepackage{lscape}
\usepackage{multirow}
\usepackage{ulem}
\usepackage{bm} 
\usepackage{cases}
\usepackage{csquotes}
\usepackage[usenames,dvipsnames]{xcolor}
\newcommand{\fig}{Fig.}

\title[DEUS-PUR Matter Power Spectrum Covariance Matrix]{Matter Power Spectrum Covariance Matrix from the DEUS-PUR $\Lambda$CDM simulations: Mass Resolution and non-Gaussian Errors}
\author[Blot et al.]{L. Blot$^1$\thanks{linda.blot@obspm.fr}, P.S. Corasaniti$^1$, J.-M. Alimi$^{1,2}$, V. Reverdy$^1$, Y. Rasera$^1$\\ 
$^1$Laboratoire Univers et Th\'eories, UMR 8102 CNRS, Observatoire de Paris, Universit\'e Paris Diderot, 5 Place Jules Janssen, 92190 Meudon, France\\
$^2$Institut d{'}Astrophysique, UMR 7095 CNRS, Universit\'e Pierre et Marie Curie, 98bis Blvd Arago, 75014 Paris, France}

\begin{document}

\maketitle
\begin{abstract}
The upcoming generation of galaxy surveys will probe the distribution of matter in the universe with unprecedented accuracy. Measurements of the matter power spectrum at different scales and redshifts will provide stringent constraints on the cosmological parameters. However, on non-linear scales this will require an accurate evaluation of the covariance matrix. Here, we compute the covariance matrix of the 3D matter density power spectrum for the concordance $\Lambda$CDM cosmology from an ensemble of N-body simulations of the Dark Energy Universe Simulation - Parallel Universe Runs (DEUS-PUR). This consists of 12288 realisations of a $(656\,h^{-1}\,\textrm{Mpc})^3$ simulation box with $256^3$ particles. We combine this set with an auxiliary sample of 96 simulations of the same volume with $1024^3$ particles. We find N-body mass resolution effect to be an important source of systematic errors on the covariance at high redshift and small intermediate scales. We correct for this effect by introducing an empirical statistical method which provide an accurate determination of the covariance matrix over a wide range of scales including the Baryon Oscillations interval. Contrary to previous studies that used smaller N-body ensembles, we find the power spectrum distribution to significantly deviate from expectations of a Gaussian random density field at $k\gtrsim 0.25\,h\,\textrm{Mpc}^{-1}$ and $z<0.5$. This suggests that in the case of finite volume surveys an unbiased estimate of the ensemble averaged band power at these scales and redshifts may require a careful assessment of non-Gaussian errors more than previously considered.
\end{abstract}


\section{Introduction}\label{intro}
Surveys of the large scale structures have been providing insightful data for more than a decade now. Observational projects such as the 2-degree Field Galaxy Redshift Survey \citep[2dFGRS,][]{Percival2001,Cole2005} and the Sloan Digital Sky Survey \citep[SDSS,][]{Tegmark2004} have yielded unprecedented measurements of the clustering of matter on the large scales. These observations have made possible the first detection of the Baryon Acoustic Oscillations (BAO) signal \citep{Eisenstein2005,Percival2007} and provided constraints on model parameters that are complementary to those obtained from other standard cosmological probes. The success of these projects has opened the way to a new generation of survey programs.

In the years to come the Dark Energy Survey\footnote{www.darkenergysurvey.org} (DES), the Large Synoptic Survey Telescope\footnote{www.lsst.org} (LSST) or the Euclid mission\footnote{www.euclid-ec.org} will map the distribution of galaxies in larger cosmic volumes and with higher sensitivity. Through a variety of probes, these surveys aim to achieve a few percent error on the determination of several cosmological parameters. However, multiple challenges need to be addressed. From the theoretical point of view one of the most challenging aspects concerns the availability of robust theoretical predictions of the clustering of matter at small scales, which is crucial to correctly interpret the data and infer unbiased constraints on model parameters. 

At small scales and late times the gravitational collapse becomes a highly non-linear process. Thus, model predictions cannot rely on standard linear perturbation theory and require solving the complex dynamics of matter collapse through numerical simulations. The need for accurate predictions of the clustering of matter over large interval of scales has driven up the demand for large volume high-resolution N-body simulations \citep{MXXL,HORIZONRUN,alimi12,darkSky}. As an example the Baryon Oscillation Spectroscopic Survey \citep[BOSS,][]{Dawson2013} has recently determined the cosmic distance scale to one percent accuracy from measurements of the BAO spectrum in the range of modes $0.01<k[\,h\,\textrm{Mpc}^{-1}]<0.30$ \citep{Anderson2013}. Future surveys such as LSST or Euclid will push these measurements even further by probing the clustering of matter up to $k\approx 1 \,h\,\textrm{Mpc}^{-1}$ \citep[see e.g.][]{Abell2009,Amendola2013}. Predicting the matter power spectrum in such a large interval of scales at a few percent level is a very challenging task since it requires N-body simulations that cover very large cosmic volumes to reduce cosmic variance uncertainties at low $k$ and with sufficient mass and spatial resolution to limit numerical systematic errors at large $k$. To tackle this challenge \citet{alimi12} have realised a series of N-body simulations, the Dark Energy Universe Simulation Full Universe Runs (DEUS-FUR), covering a cosmic volume of $(21\,\textrm{Gpc}\,h^{-1})^3$ with $8192^3$ particles which has allowed to resolve the BAO spectrum to $1\%$ accuracy level \citep[][]{Rasera2014}. However, this is still not sufficient to correctly interpret the data since an unbiased statistical analysis also require knowledge of the covariance matrix. Addressing this issue is the goal of the analysis presented here.
 
If the initial matter density field is Gaussian distributed, during the linear regime its Fourier modes evolve independently and the covariance of the matter power spectrum has a simple diagonal form. From now on we will refer to this configuration as the Gaussian case, for which errors on the band powers are uncorrelated. In contrast at small scales and late times the covariance develops non-vanishing off-diagonal terms which account for the mode coupling caused by the non-linear regime of gravitational collapse. In such a case the errors on band powers become correlated causing a larger dispersion on power spectrum measurements \citep{meiksin99}. Neglecting such correlations may lead to biased results as shown by several studies of weak lensing observables \citep[see e.g.][]{WhiteHu00,Semboloni07,LeePen08,Kiessling11} and to a biased determination of BAO parameters \citep[see e.g.][]{takahashi11,Ngan12}. Hence, the future generation of large scale structure surveys will need estimates of the covariance matrix which require sampling the matter power spectrum from large samples of N-body simulations \citep[][]{taylor13}. 

We undertake this task using a set of simulations $(\sim10^4)$ of the concordance $\Lambda$CDM model from the Dark Energy Universe Simulation Parallel Universe Runs (DEUS-PUR). By combining this large ensemble with an auxiliary set of high resolution runs we estimate the impact of the mass resolution of the simulations on the 3D matter density power spectrum covariance matrix. We show that mass resolution effects can be an important source of uncertainty. Furthermore, our analysis clearly indicates the necessity of using large N-body datasets when studying the matter power spectrum distribution at small scales and low redshifts. In particular we find significant deviations from expectations of a Gaussian density field which were overlooked in previous studies. This has potentially important implications for inferring unbiased power spectrum measurements for future survey programs. Our work extends the analysis of \citet{takahashi09} to a larger statistical sample of N-body simulations with higher mass resolution and better spatial resolution.

The paper is organised as follows: in Section \ref{nbodydata} we describe the characteristics of the DEUS-PUR simulations. In Section \ref{deuspurcov} we present the analysis of the DEUS-PUR covariance matrix, while in Section \ref{pdfsec} we show the results of the computation of the probability distribution of the matter power spectrum. In Section \ref{sn} we discuss the effect of non-Gaussian errors on the signal-to-noise of power spectrum measurement. Finally we present our conclusion in Section \ref{conclu}.

\section{N-body Dataset}\label{nbodydata} 
\subsection{DEUS-PUR Simulations}
We use the N-body simulation dataset from DEUS-PUR project. This consists of 12288 simulations of a flat $\Lambda$CDM model with parameters (see Table \ref{tab:cosmopar}) calibrated to the WMAP-7yr data~\citep{spergel07} of a cosmological volume of $(656\,h^{-1}\,\textrm{Mpc})^3$ with $256^3$ particles, for a formal mass resolution of $1.2\times 10^{12}\,h^{-1}\,\textrm{M}_\odot$ (Set A), and 96 simulations of the same cosmological model and equal volume with $1024^3$ particles, corresponding to a mass resolution of $2\times 10^{10}\,h^{-1}\,\textrm{M}_\odot$ (Set B). These runs have been realised with ``A Multiple purpose Application for Dark Energy Universe Simulation" (AMADEUS) \citep{alimi12}. This workflow application includes a dynamical solver based on RAMSES~\citep{teyssier02}, an adaptive mesh refinement code with a tree-based data structure that allows recursive grid refinement on a cell-by-cell basis, in which particles are evolved using a particle-mesh (PM) solver, while the Poisson equation is solved using a multigrid method \citep{guilletteyssier11}. In the case of the DEUS-PUR simulations the refinement criterion is set such as to allow up to 6 levels of refinement.

\begin{table}
\centering
\begin{tabular}{c|c|c|c|c}
\hline
\ $h$ & $\Omega_m h^2$ & $\Omega_b h^2$ & $n_s$ & $\sigma_8$ \\
\hline
\ $0.72$ & $0.1334$ & $0.02258$ & $0.963$ & $0.801$ \\
\hline
\end{tabular}
\caption{DEUS-PUR cosmological model parameter values.}\label{tab:cosmopar}
\end{table}

The initial conditions of the simulations have been generated using the code MPGRAFIC~\citep{prunet08}, which convolves a white noise with the square root of the input power spectrum and generates initial displacements and velocities of the dark matter particles using the Zeldovich approximation. In order to avoid the generation of $\sim 10^4$ white noises we generated $3$ independent white noises in $3$ cubes of $4096^3$ particles each that we have subsequently split into $4096$ sub-cubes. The latter were then used separately to generate the initial conditions of each simulation of Set A. A similar procedure has been used to generate the initial conditions of the simulations of Set B. These have been obtained by splitting $2$ different white noises of $4096^3$ particles in $64$ cubes. The initial redshift has been set such that all simulations starts with the same amplitude of density fluctuations at the scale of the grid resolution. This is a standard technique that allows to consistently compare simulations with different spatial and mass resolution. Let us denote by $\sigma(L/N_p,z_i)$ the root-mean-square fluctuation of the linear density field at an initial redshift $z_i$ on the scale of the grid resolution $L/N_p$, where $L$ is the simulation box-length and $N_p$ is the number of particles. Then, $z_i$ is determined by setting $\sigma(L/N_p,z_i)$ to an arbitrary small value and iteratively solving the algebraic equation
\begin{equation}
\frac{\sigma(L/N_p,z_i)}{\sigma(L/N_p,z=0)}=\frac{D_+(z_i)}{D+(z=0)},
\end{equation} 
where $D_+$ is the linear growth function. We set $\sigma(L/N_p,z_i)=0.02$, such that initial redshifts are sufficiently large to ensure the validity of the Zeldovich approximation, but not exceedingly large such as to avoid the introduction of systematic effects due to integration of numerical noise. In the case of simulation Set A this gives $z_i\approx 105$, while for the higher resolution Set B we have $z_i\approx190$. Such large values guarantee that transient effects \citep{Scoccimarro98,Crocce06} are negligible. Table~\ref{tab:sims} summarises the characteristics of the DEUS-PUR simulations. 

\citet{takahashi09} have performed a study of the 3D power spectrum covariance matrix from 5000 simulations of a standard $\Lambda$CDM model with a $(1000\,h^{-1}\,\textrm{Mpc})^3$ volume and a mass resolution of $4.1\times 10^{12}\,h^{-1}\,\textrm{M}_\odot$, realised with a Particle Mesh (PM) solver with no spatial refinement and initial redshift $z_i=20$. For comparison our Set A has nearly $3$ times more realisations which allow us to determine the covariance matrix with reduced statistical errors. Furthermore, the DEUS-PUR simulations have a mass resolution $\sim 3$ times higher, a better spatial resolution at the level of the coarse grid by nearly a factor of $1.5$ (and a factor of $\sim100$ at the most refined level) and start at a much higher redshift, thus allowing us to reduce the effect of numerical systematics when compared to the sample used by \citet{takahashi09}.

The workflow of the AMADEUS application has been automated to generate a large number of N-body simulations. An external script has been coded to monitor in real time the job-queue and submit new simulations as soon as other simulations have terminated. For each simulation the initial conditions, the dynamic evolution, the data reduction and measurements of the matter power spectrum and the halo mass function are controlled through the same script. A final check on the file content has been implemented to detect any error due to unexpected machine failure. Each AMADEUS script has been launched as a separate job on the ADA supercomputer \footnote{http://www.idris.fr/eng/ada/hw-ada-eng.html} of the Institute for Development and Resources in Intensive Scientific Computing (IDRIS). Simulations of Set A were run on 8 processors Intel Sandy Bridge E5-4650 for a running time of $\sim1$ hour per simulation, while Set B simulations took $\sim24$ hours per simulation on 64 processors.

\begin{table}
\centering
\begin{tabular}{c|c|c|c|c}
\hline
\ Set & $N_s$ & $L$ ($\textrm{Mpc}\;h^{-1}$) & $N_p$ & $m_p$ ($M_\odot\,h^{-1}$) \\
\hline
\ A & $12288$ & $656.25$ & $256^3$ & $1.2 \times 10^{12}$ \\
\ B & $96$ & $656.25$ & $1024^3$ & $2 \times 10^{10}$ \\
\hline
\end{tabular}
\caption{DEUS-PUR simulation characteristics: $N_s$ is the number of realisations, $L$ is the box-side length, $N_p$ is the number of dark matter particles and $m_p$ the mass resolution. Taking set A as a reference, set B has been designed to study simulation mass resolution effects on large scale structure observables.}
\label{tab:sims}
\end{table}

\subsection{Power Spectrum \& Covariance Matrix Estimators}
We compute the matter power spectrum using the code POWERGRID ~\citep{prunet08}. This estimates the power spectrum from the Fourier transform of the matter density field in band powers of size $\Delta k=2\pi/L$. We correct the measured spectrum for the effect of smoothing due to the Could-In-Cell (CIC) algorithm, that is used to estimate the density contrast field from the particle distribution. We do not correct for aliasing, since varying the size of the CIC grid we find that aliasing effects are negligible below half the Nyquist frequency of the CIC grid. Since our CIC grid is two times finer than the coarse grid of the simulation, its Nyquist frequency is given by $k_{N}=2\,(\sqrt[3]{N_p}\,\pi/L)$. Thus the range of modes in which we compute the power spectrum is given by $k_{\textrm{min}}=2 \pi /L$ and $k_{\textrm{max}}=k_{N}/2$. More specifically $k_{\textrm{min}}\approx 0.01\,h\,\textrm{Mpc}^{-1}$ for both sets A and B, while $k_{\textrm{max}}\approx 1.22\,h\,\textrm{Mpc}^{-1}$ for set A and $k_{\textrm{max}}\approx 5.9\,h\,\textrm{Mpc}^{-1}$ for set B. To be conservative we restrict our analysis to Fourier modes up to $k\approx1\,h\,\textrm{Mpc}^{-1}$.

The covariance matrix is computed using the unbiased sample covariance estimator:
\begin{equation}\label{cov_def}
\widehat{\text{cov}}(k_1,k_2) = \frac{1}{N_s-1}\sum_{i=1}^{N_s}[\hat{P}_i(k_1)-\bar{P}(k_1)][\hat{P}_i(k_2)-\bar{P}(k_2)],
\end{equation}
where $N_s$ is the number of independent realisations and $\bar{P}(k)=\sum_{i=1}^{N_s}\hat{P}_i(k)/N_s$ is the sample mean, with $\hat{P}_i(k)$ the matter power spectrum estimation of the $i$-th realisation.

\section{DEUS-PUR Covariance Matrix}\label{deuspurcov}
Let us consider the formal expression of the matter power spectrum covariance matrix \citep[see e.g.][]{Scoccimarro1999}:
\begin{equation}\label{covdev}
\begin{split}
\text{cov}(k_1,k_2)&=\frac{2}{N_{k_1}}P^2(k_1)\delta_{k_1,k_2} +\\
&+\frac{1}{V} \int_{\Delta_{k_1}} \int_{\Delta_{k_2}} \frac{d^3\mathbf{k^\prime_1}}{V_{k_1}} \frac{d^3\mathbf{k^\prime_2}}{V_{k_2}} T(\mathbf{k^\prime_1}, -\mathbf{k^\prime_1}, \mathbf{k^\prime_2}, -\mathbf{k^\prime_2}),
\end{split}
\end{equation}
where $P(k)$ is the matter power spectrum, $N_k\approx k^2\Delta k \,V/(2\pi^2)$ is the number of $k$-modes in the volume $V$, $\Delta_{k_i}$ is the band power integration interval centred on the mode $k_i$ and $V_{k_i}$ is the integration volume in Fourier space; the integrand $T(\mathbf{k^\prime_1}, \mathbf{k^\prime_2}, \mathbf{k^\prime_3}, \mathbf{k^\prime_4})$ is the trispectrum (the fourth order connected moment in Fourier space) of the density fluctuation field. The first term in Eq.~(\ref{covdev}) represents the Gaussian contribution to the covariance. As already mentioned, for an initial Gaussian density field, during the linear regime the Fourier modes evolve independently and the power spectrum covariance is diagonal with amplitude $2 P^2(k)/N_k$. The second term in Eq.~(\ref{covdev}) represents the contribution of non-Gaussianity arising during the non-linear regime of gravitational collapse at small scales. 

Non-linearities induce mode couplings which source a non-vanishing trispectrum of the density fluctuation field. Since this cannot be computed exactly then the covariance matrix must be estimated by sampling the matter power spectrum from a large ensemble of numerical N-body simulations. This computation is not exempt of systematic uncertainties. For instance, the finite volume of simulations is source of non-Gaussian errors \citep{Rimes2006} and as shown in \citet{takahashi09} this can introduce large uncertainties even on weakly non-linear scales. We leave a detailed study of this effect to a forthcoming work. In the following, we focus on systematic errors due to the mass resolution of the simulations, which have been neglected in previous studies.

\begin{figure}
\begin{center}
\includegraphics[scale=0.5]{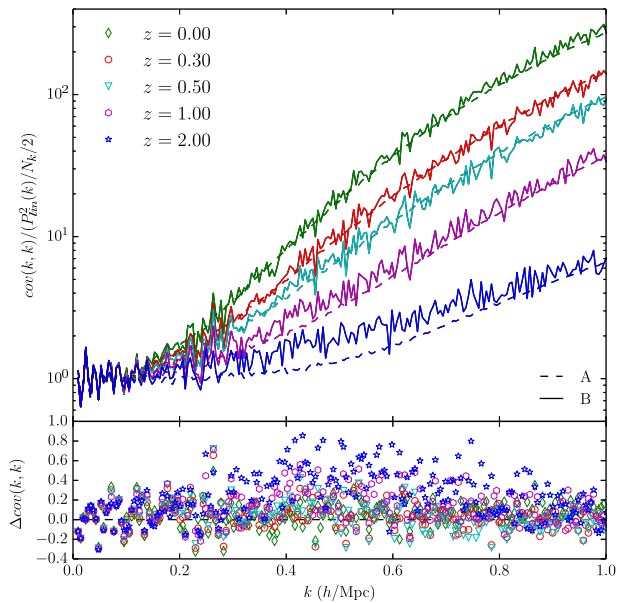}
\caption{Top panel: diagonal elements of the covariance matrix normalised to the Gaussian variance for Set A (dashed lines) and B (solid lines) at $z=0$ (green), $0.3$ (red), $0.5$ (light-blue), $1$ (magenta) and $2$ (blue) respectively. Bottom panel: relative difference between Set A and B at different redshifts. The variance of Set A is underestimated compared to that of Set B.}
\label{variance}
\end{center}
\end{figure}

\begin{figure}
\begin{center}
\includegraphics[scale=0.5]{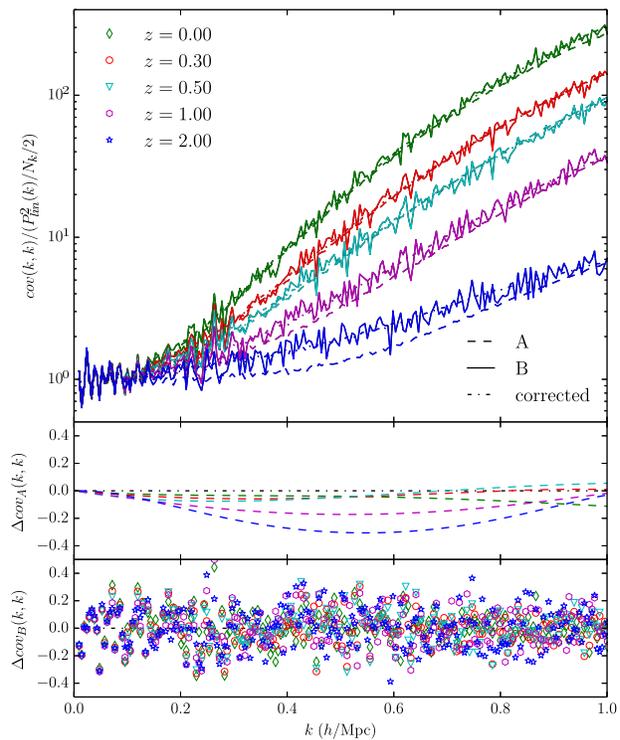}
\caption{Top panel: as in Fig.~\ref{variance} including the diagonal components of the covariance matrix from the corrected spectra of Set A (dot-dashed line). Middle panel: relative difference of the uncorrected variance of Set A (central) with respect to the corrected one. Corrections can be as large as $40$ per cent at $z=2$, $20$ per cent at $z=1$ and less than $10$ per cent at lower redshifts. Bottom panel: relative difference of the variance from the higher resolution simulations Set B with respect to the correct variance. The residuals show no systematic shift indicating that the correction efficiently accounts for the mass resolution effect.}
\label{variance_c}
\end{center}
\end{figure}

\subsection{Numerical Simulation Mass Resolution Errors}\label{massres}
In the top panel of Fig.~\ref{variance} we plot the diagonal elements of the matter power spectrum covariance matrix normalised to the linear Gaussian amplitude $2\,P^2_{lin}(k)/N_k$ for Set A (dash lines) and B (solid lines) at $z=0,0.3,0.5,1$ and $2$ (top to bottom) respectively. As we can see the curves corresponding to Set A are very smooth since they have negligible noise due to the large size of the simulation sample. This is not the case of Set B for which the covariance estimates are characterised by a higher level of noise. As expected, the onset of the non-linear regime causes deviations from the Gaussian prediction which occur at large $k$-values and shift towards smaller ones at lower redshifts. For instance, a deviation of a factor $\sim 5$ at $z=1$ occurs at $k\sim 0.55\,h\,\textrm{Mpc}^{-1}$, while at $z=0$ the same deviation occurs at $k\sim 0.30\,h\,\textrm{Mpc}^{-1}$. The effect of such deviations is to increase the statistical errors on the power spectrum measurements at non-linear scales. Despite the higher level of statistical noise associated to Set B, it is evident that there is a systematic down shift of the variance of lower resolution simulations. In the bottom panel of Fig.~\ref{variance} we can see that such a discrepancy exceeds the statistical noise of Set B at redshifts $z>0.5$ in the range of modes $0.20\lesssim k[\,h\,\textrm{Mpc}^{-1}]\lesssim 0.80$ with an amplitude that on average can be as large as $\sim 40$ per cent. This is a direct consequence of the suppression of the matter power spectrum at large/intermediate $k$ for lower mass resolution simulations.
  
The mass resolution of numerical simulations is a known source of systematic errors and a generic feature of simulations which rely on the PM method~\citep{joyce09,heitmann10,Rasera2014}. The suppression of power at small scales for lower mass resolution simulations is due to the combination of the precision of force calculation, which is based on grid discretization of the gravitational field, and the particle sampling of the matter density field. In a PM code the number of coarse grids is usually set equal to the number of particles, therefore, before any refinement is triggered, particles in under-dense or over-dense regions will experience a reduced force by the close environment compared to higher resolution simulations. \citet{knebe01} showed that for PM based codes the ideal configuration is to have $8$ times more grid points than particles, but in a cosmological simulation this requirement may be too expensive from the computational point of view. None the less, the artificial suppression of power is mitigated at low redshifts and/or higher $k$ when the local density of particles in the simulations triggers the AMR grid refinement, which explains why this systematic effect shown in Fig.~\ref{variance} fades away across the whole interval at $z\le 0.5$.

\citet{Rasera2014} corrected the BAO spectrum for the mass resolution effect by combining the cosmic variance limited power spectrum from DEUS-FUR with that of smaller volume and higher resolution simulations. Here, we opt for a similar strategy. In fact, the covariance is obtained by sampling the matter power spectrum of independent realisations, thus we can correct the lower resolution power spectra of Set A by implementing statistical information on the power spectra obtained from the higher resolution simulations of Set B. In Appendix \ref{appendix} we provide a detailed derivation of the correction, which we assume to be linear:
\begin{equation}\label{pcorr}
\hat{P}^{\rm corr}_{\textrm{A}}(k) = \left[\hat{P}_{\textrm{A}}(k)-\bar{P}_{\textrm{A}}(k)\right] \frac{\sigma_{\hat{P}_{\textrm{B}}}(k)}{\sigma_{\hat{P}_{\textrm{A}}}(k)}  + \bar{P}_{\textrm{B}}(k),
\end{equation}
where $\bar{P_{\alpha}}(k)$ and $\sigma_{P_{\alpha}}(k)$ ($\alpha=\textrm{A},\textrm{B}$) are the average and root-mean-square of the power spectrum estimator respectively.

In \fig~\ref{variance_c} we show the diagonal elements of the covariance matrix normalised to the Gaussian term for Set A before (dash lines) and after (dash dot lines) correction, and for set B (solid lines) at $z=0.0,0.3,0.5,1.0$ and $2.0$ (top to bottom) respectively. As we can see the corrected curves fit well through those obtained from the higher resolution simulations. In the middle panel of \fig~\ref{variance_c} we show the relative differences between the corrected and uncorrected curves at different redshifts (dash lines top to bottom corresponds to $z=0$ to $2$). We can clearly see that the amplitude of the correction at high redshift in the interval $0.20\lesssim k[\,h\,\textrm{Mpc}^{-1}]\lesssim 0.80$ can be as large as $40$ per cent, while at $z<0.5$ the difference remains below the $10$ per cent level. In the bottom panel of \fig~\ref{variance_c} we also plot the residual between the corrected covariance of Set A and that from Set B (different dot types correspond to different redshifts as shown in the legend). As we can see there is no systematic shift and the only differences are due to the statistical noise of Set B.

The mass resolution error also underestimate the off-diagonal components of the covariance matrix, none the less the amplitude results to be smaller than the systematic shift we have seen on the diagonal elements. As expected we find the correction of the low resolution power spectra to also account for the mass resolution effect on the off-diagonal components.

\begin{figure*}
\begin{minipage}{18cm}
\begin{center}
  \includegraphics[scale=0.4]{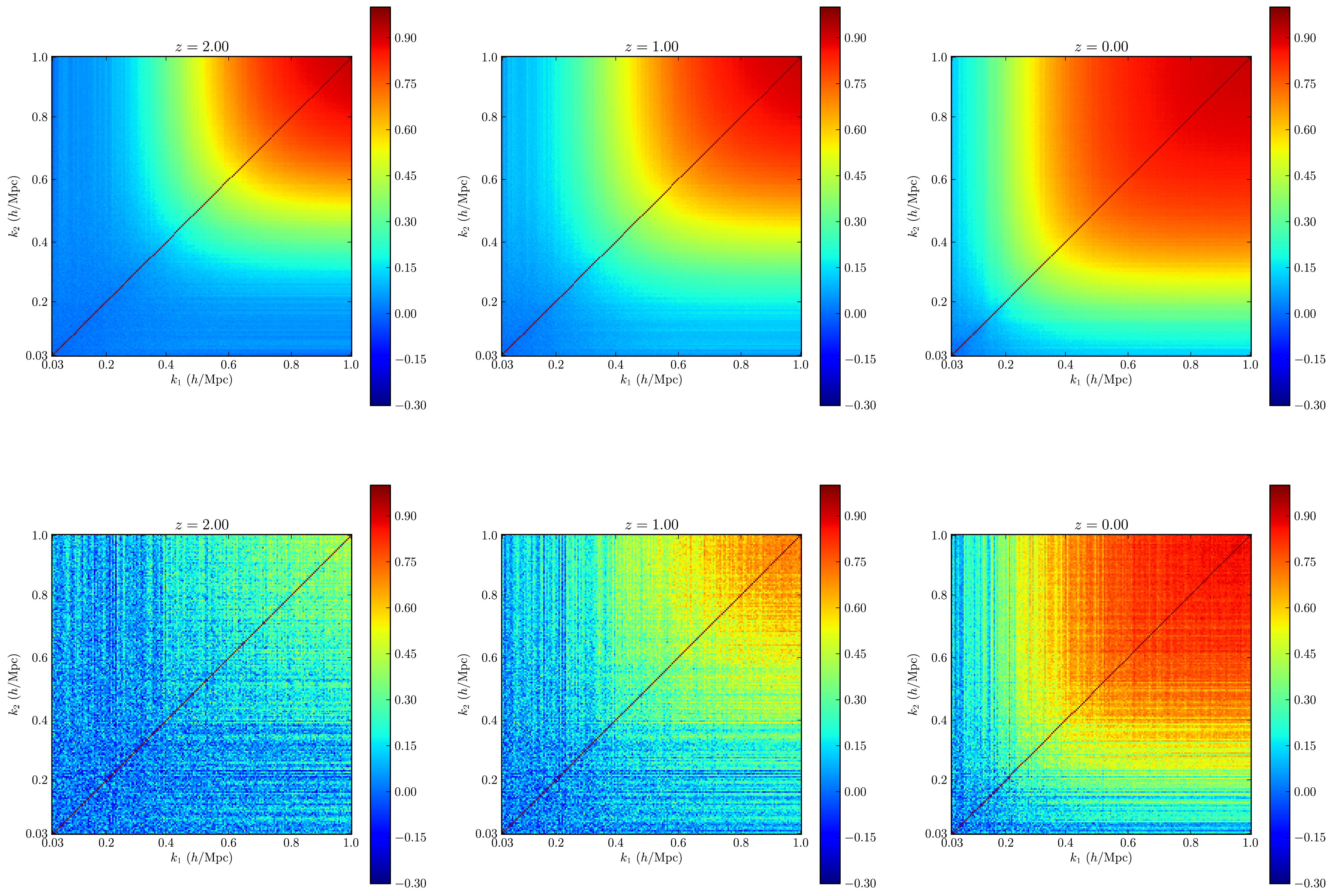}
  \caption{\label{corr_coeff} Correlation coefficient matrix at $z=2$ (left panels), $1$ (central panels) and $0$ (right panels) respectively estimated from Set A (top panels) and Set B (bottom panels). We can see the increasing amplitude of pair correlation at high $k$ shifting towards lower wave-numbers for decreasing redshift. The comparison between the two sets shows that the structure of the correlations is poorly reproduced when using a low number of simulations.}
\end{center}
\end{minipage}
\end{figure*}

\subsection{Fourier Mode Correlations}
In order to quantify the correlation between pairs of Fourier modes it is useful to introduce the correlation coefficient
\begin{equation}
r(k_1,k_2) = \frac{\text{cov}(k_1,k_2)}{\sqrt{\text{cov}(k_1,k_1)\, \text{cov}(k_2,k_2)}},
\end{equation}
which varies between $1$ (maximum correlation) and $-1$ (maximum anti-correlation), and is $0$ when modes are uncorrelated. In linear regime the correlation coefficient is the identity matrix. 

In Fig.~\ref{corr_coeff} we plot $r(k_1,k_2)$ in the interval $0.03<k[\,h\,\textrm{Mpc}^{-1}]<1.00$ (which includes the BAO range) at $z=2$ (left panels), $1$ (central panels) and $0$ (right panels) for Set A (top panels) and for Set B (bottom panels). We do not show the correlation coefficient inferred from the corrected spectra of Set A since this coincides with the uncorrected one to very good approximation. This is because the mass resolution effect discussed in the previous section scales approximately linearly with the power spectrum affecting the covariance matrix amplitude. Since correlation coefficient is given by the covariance matrix normalised by the root-square of the product of its diagonal elements the effect cancels out in the ratio.

Non-vanishing off-diagonal elements are already present at $z=2$ at large $k$ values, for instance the mode $k_1\sim 0.3\,h\,\textrm{Mpc}^{-1}$ has a $10$ per cent correlation with $k_2\sim 0.4\,h\,\textrm{Mpc}^{-1}$ and $20$ per cent with $k_2\sim 1\,h\,\textrm{Mpc}^{-1}$. The amplitude of the correlations increases as function of $k$ and extends towards smaller $k$ values at lower redshifts as the dynamics of the modes increasingly deviates from the linear regime of collapse. The comparison between the two sets shows the importance of having a large set of simulations, in order to reduce the impact of noise. In fact, the structure of correlations is much clearer for Set A than for Set B, for which at $z=2$ the signal is hard to distinguish from the statistical noise. It is worth noticing that in the BAO range ($0.01<k[\,h\,\textrm{Mpc}^{-1}]<0.30$) the correlation in the off-diagonal components can reach a level up to $30$-$35$ per cent between redshift $1$ and $0$, which confirms the need of an accurate estimation of the 3D power spectrum covariance matrix for BAO data analyses.

\section{Probability Distribution of the Power Spectrum}\label{pdfsec}
We now focus on the probability distribution function (PDF) of the matter power spectrum estimator.

In the case of Gaussian initial conditions, during the linear regime of gravitational collapse the matter power spectrum at a given wave-number $k$ is distributed as a $\chi^2$ with $N_k$ degrees-of-freedom \citep[see e.g.][]{fisher93}. In the large $N_k$ limit, which corresponds to sufficiently large volumes and high wave-numbers, the PDF tends to a Gaussian. However, at high $k$ the non-linear evolution of matter clustering is expected to introduce non-Gaussianities (i.e. departures from a $\chi^2$-distribution in the large $N_k$ limit). 

The large sample of simulations from DEUS-PUR allows us to finely sample such a distribution and test for non-Gaussianities. To this end it is convenient to rescale the power spectrum estimator as $\sqrt{N_k/2}(\hat{P}/\bar{P}-1)$, such that in the large $N_k$ limit and in linear regime the distribution is a Gaussian with zero mean and unity variance. 

\begin{figure}
\begin{center}
\includegraphics[scale=0.5]{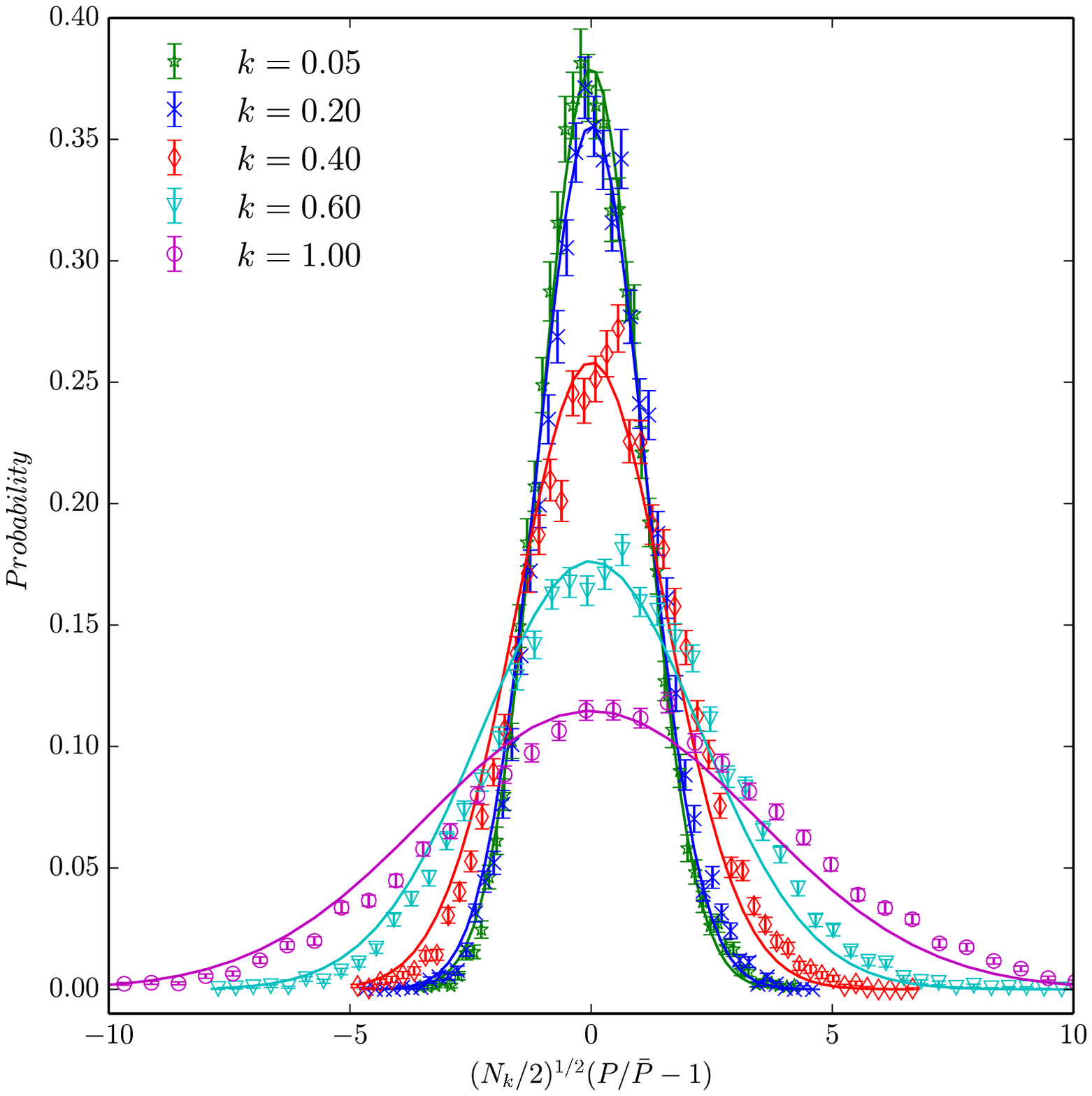}
\caption{Probability distribution of the rescaled power spectrum estimator $\sqrt{N_k/2}(\hat{P}/\bar{P}-1)$ estimated from the 12288 realisations of Set A for $k=0.05$ (green star) $0.20$ (blue cross) $0.40$ (red diamond), $0.60$ (light-blue triangle) and $1.00\,h\,\textrm{Mpc}^{-1}$ (magenta circle) respectively. The error bars are given by Poisson errors. The solid line curves show the Gaussian distribution with sample mean and variance of the same set at the corresponding values of $k$.}
\label{pdf}
\end{center}
\end{figure}

In Fig. \ref{pdf} we plot the estimated PDF from Set A at $z=0$ and $k=0.05,0.20,0.40$, $0.60$ and $1.00\,h\,\textrm{Mpc}^{-1}$ respectively. The error bars are given by Poisson errors. At $k\le 0.20\,h\,\textrm{Mpc}^{-1}$ we can see that the PDF is consistent with a Gaussian distribution within statistical errors, while at higher $k$ we can clearly see deviations from Gaussianity above statistical uncertainties. We quantify these deviations in terms of the skewness and kurtosis defined as:
\begin{align}
S_3(k) &= \frac{N_s^{-1} \sum_{i=1}^{N_s}[\hat{P}_i(k) - \bar{P}(k)]^3}{\big\{N_s^{-1} \sum_{i=1}^{N_s}[\hat{P}_i(k) - \bar{P}(k)]^2 \big\}^ {3/2}},\label{s3}\\
S_4(k) &= \frac{N_s^{-1} \sum_{i=1}^{N_s}[\hat{P}_i(k) - \bar{P}(k)]^4}{\big\{N_s^{-1} \sum_{i=1}^{N_s}[\hat{P}_i(k) - \bar{P}(k)]^2 \big\}^ {2}}-3.\label{s4}
\end{align}
In the case of a $\chi^2$-distribution with $N_k$ degree-of-freedom these can be computed exactly resulting in $S_3(k)=\sqrt{8/N_k}$ and $S_4(k)=12/N_k$ \citep[see e.g.][]{takahashi09}. 

\begin{figure}
\begin{center}
\includegraphics[scale=0.5]{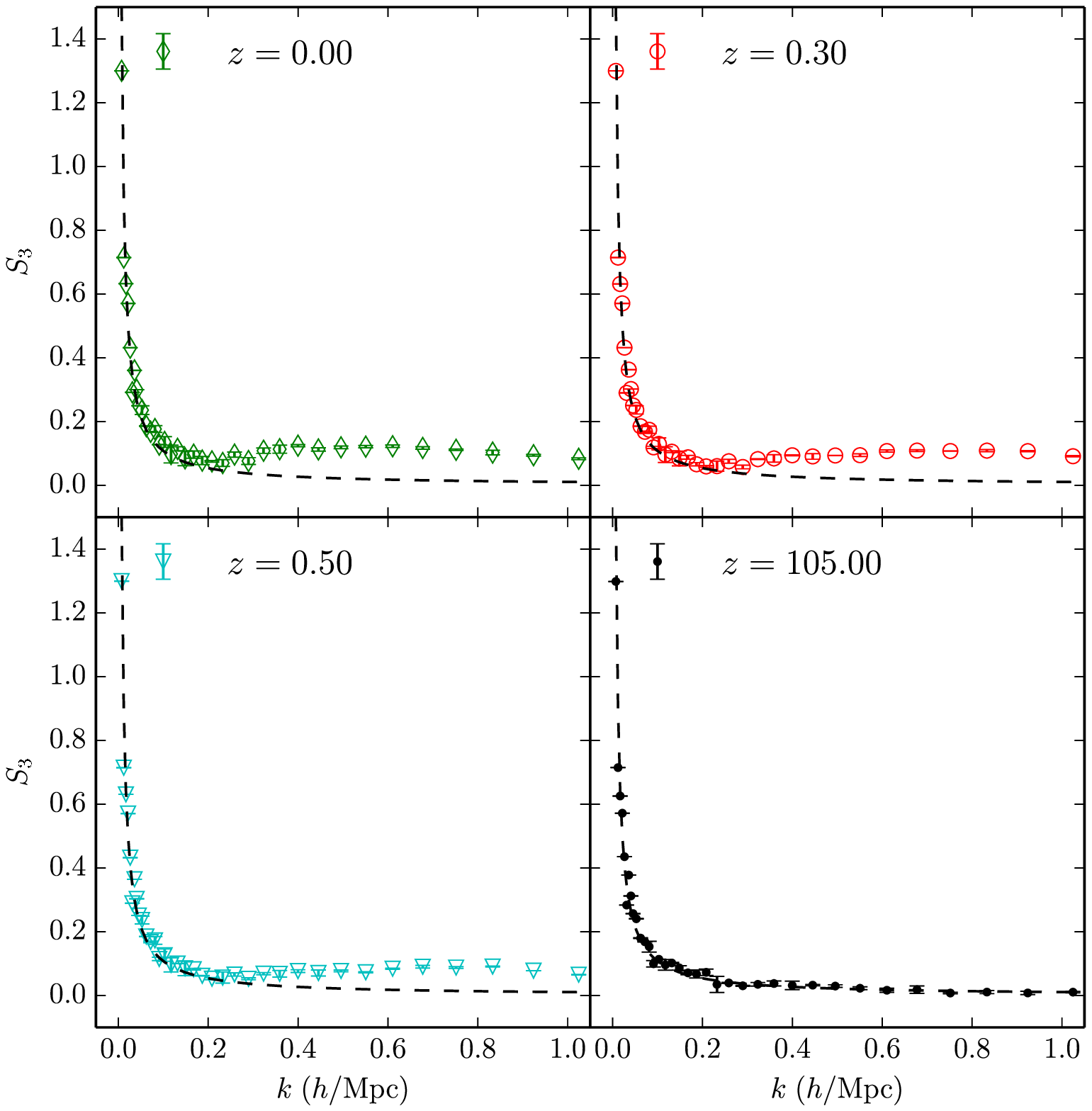}
\includegraphics[scale=0.5]{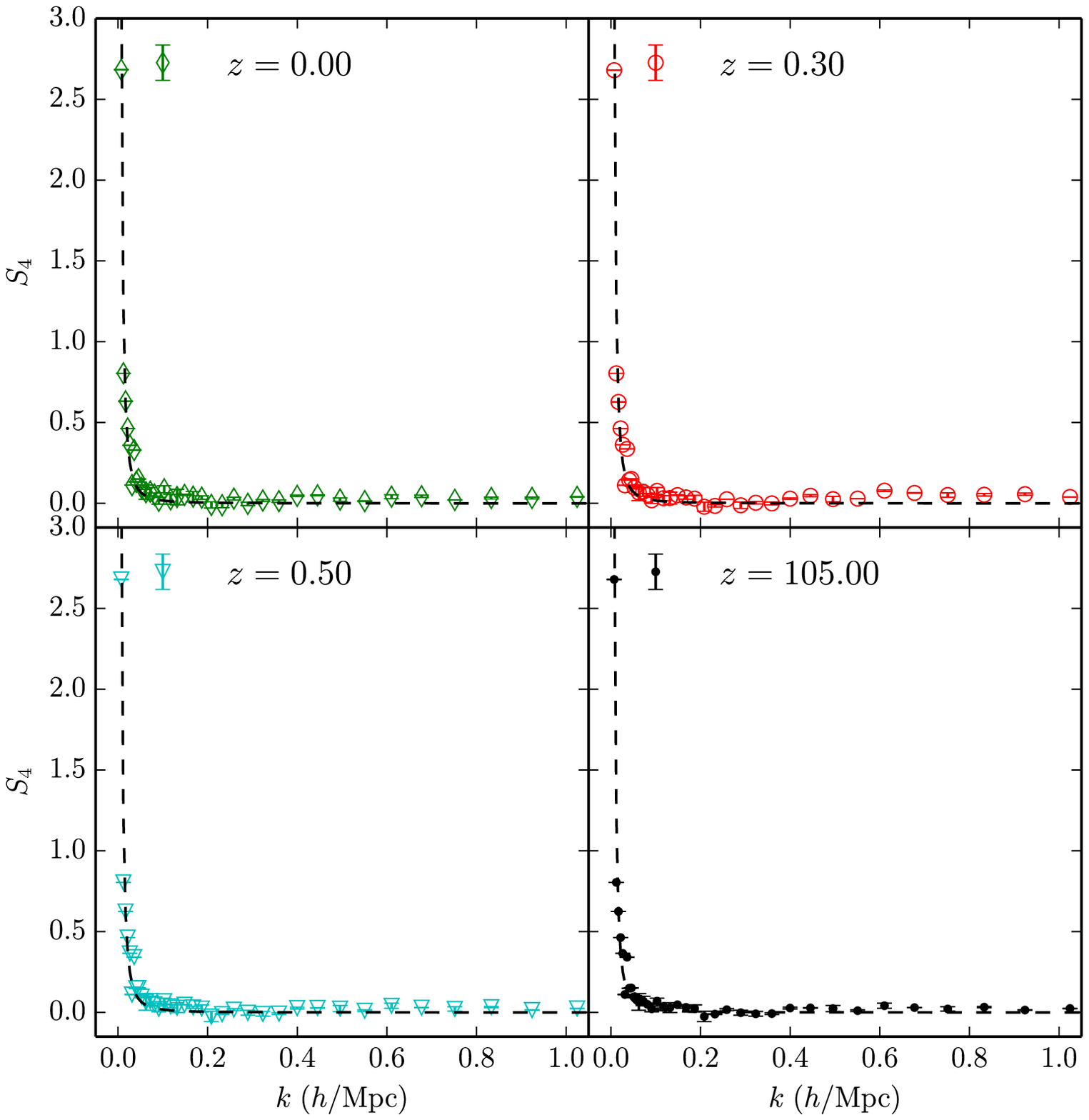}
\caption{Skewness (top panels) and kurtosis (bottom panels) of the probability distribution of $\hat{P}$ as a function of $k$ for $z=0,0.3,0.5$ and $105$ respectively. The measured values are binned in intervals of size $\Delta k/k\sim0.1$, the associated Poisson errors are smaller than the data points. The dashed lines represent the $\chi^2$-distribution predictions.}
\label{high}
\end{center}
\end{figure}

In Fig.~\ref{high} we plot $S_3(k)$ (top panels) and $S_4(k)$ (bottom panels) from Set A estimated at $z=105$ and $z=0,0.3,0.5$ (where mass resolution effects are subdominant) respectively. For visual purposes we have binned the estimated values in bins of size $\Delta k/k=0.1$ and included statistical errors on the data points. The dashed lines corresponds to the $\chi^2$ expected values of $S_3(k)$ and $S_4(k)$. We notice that at $z=105$ the skewness is consistent with that from the $\chi^2$-distribution, while for $z<0.5$ and $k\gtrsim 0.25\,h\,\textrm{Mpc}^{-1}$ we can clearly see increasing deviations as function of $k$ at high statistical significance. In contrast the kurtosis remains consistent with $\chi^2$ expected values and any departure of $S_4(k)$ from the Gaussian random field prediction still remains within statistical errors.

Previous studies have determined the power spectrum distribution using smaller simulation ensembles and at low redshifts found no statistically significant deviation of the skewness from expectations of a Gaussian random density field \citep[see e.g.][]{takahashi09}. This stresses the necessity of using very large samples of simulations. We believe that such a result can have important observational implications which warrant further investigation. At large $k$ and $z<0.5$ the ratio $S_3(k)/\sqrt{8/N_k}\gtrsim 2$, hence unbiased measurements of the band power from observables of the clustering of matter such as weak lensing observations \citep[see also][]{sato2009} may require prior knowledge of the $\hat{P}(k)$ distribution. At lower $k$ the departure from a $\chi^2$-distribution is at most of a factor $2$ for $k\lesssim 0.30\,h\,\textrm{Mpc}^{-1}$. Thus, measurements of the BAO may still be performed using only covariance matrix information, though aiming at sub-percent accuracy may require a more detailed study to elucidate the full impact of the non-Gaussian distribution of $\hat{P}(k)$.

\section{Signal-to-Noise}\label{sn}
We estimate the effect of the correlation between band power errors on the signal-to-noise of the matter power spectrum
\begin{equation}
\left(\frac{S}{N}\right)^2 = \sum_{k_1,k_2 < k_{\rm max}} P(k_1)\, \psi(k_1,k_2)\, P(k_2),\label{ston}
\end{equation}
where $\psi=\text{cov}^{-1}(k_1,k_2)$ is the inverse of covariance matrix, also known as precision matrix. Since we estimate the covariance from a finite ensemble of independent realisations, there is a statistical error associated to the sample covariance. Thus, inverting the sample covariance gives a biased estimate of the precision matrix. For a Gaussian random density field the unbiased estimator of the precision matrix is given by \citep[see e.g.][]{Hartlap2007,taylor13}: 
\begin{equation}\label{unbiasedinvcov}
\widehat{\psi}(k_1,k_2) = \frac{N_{\rm s} - N_{\rm b} -2}{N_{\rm s}-1}\, \widehat{\text{cov}}^{-1}(k_1,k_2),
\end{equation}
where $\widehat{\text{cov}}$ is the covariance estimator defined in Eq.~(\ref{cov_def}), $N_{\rm s}$ is the number of realisations and $N_{\rm b}$ is the number of band power bins. This estimator is defined only for $N_{\rm s}> N_{\rm b}+2$. In any case, for $N_{\rm s}< N_{\rm b}+1$ the values of the sample covariance are not positive definite and its inverse is not defined~\citep{Hartlap2007}.

In evaluating the signal-to-noise we set the power spectrum in Eq.~(\ref{ston}) to the average of the corrected spectra from Set A and compute the signal-to-noise using the precision matrix defined by Eq.~(\ref{unbiasedinvcov}) for the corrected and uncorrected spectra of Set A respectively. 

\begin{figure}
\begin{center}
\includegraphics[scale=0.5]{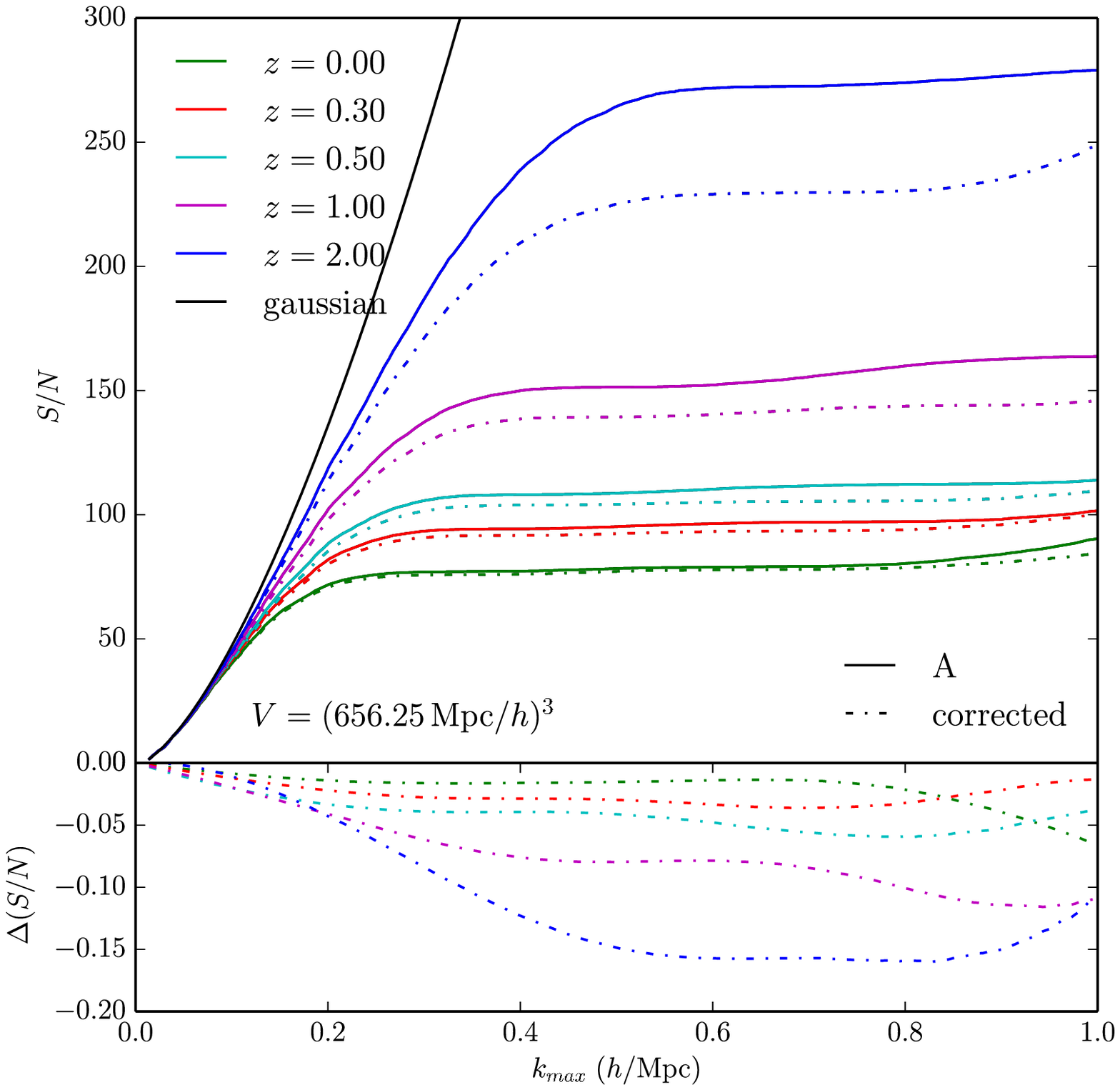}
\caption{Top panel: signal-to-noise of power spectrum measurements as function of $k_{\rm max}$ estimated from Set A with (solid line) and without (dash line) mass resolution correction for $z=0,0.3,0.5,1$ and $2$ respectively. The solid black line corresponds to the Gaussian prediction. Bottom panel:  relative difference of the S/N with and without mass resolution correction.}
\label{signoise}
\end{center}
\end{figure}

In Fig.~\ref{signoise} we plot the resulting signal-to-noise estimates at $z=0,0.3,0.5,1$ and $2$ respectively as function of $k_{\rm max}$. We can see that the effect of mass resolution errors is to artificially enhance the signal-to-noise. As discussed in Section~\ref{massres} this is because lower mass resolution simulations underestimate the covariance matrix. This results into a greater amplitude of the precision matrix components and consequently in a larger signal-to-noise compared to higher mass resolution estimates. As we can see in Fig.~\ref{signoise} the signal-to-noise from the corrected Set A is up to $\sim 15$ per cent smaller at $k_{\rm max}\gtrsim 0.30\,h\,\textrm{Mpc}^{-1}$, while at lower redshift (where the mass resolution effect is negligible) the S/N from the corrected and uncorrected Set A agree within a few percent. For comparison, we also plot the expected S/N in the Gaussian case. Assuming that the precision matrix from Set A is drawn from the inverse-Wishart distribution~\citep{Press82}, then from the study of \citet{taylor13} we expect the statistical errors on S/N to be $\sim 1$ per cent, much smaller than the effect of mass resolution at $z>1$. As already noted by \citet{Angulo2008}, the signal-to-noise saturates above a redshift dependent scale \citep[see also][]{Smith2009,takahashi09}.

\begin{figure}
\begin{center}
\includegraphics[scale=0.5]{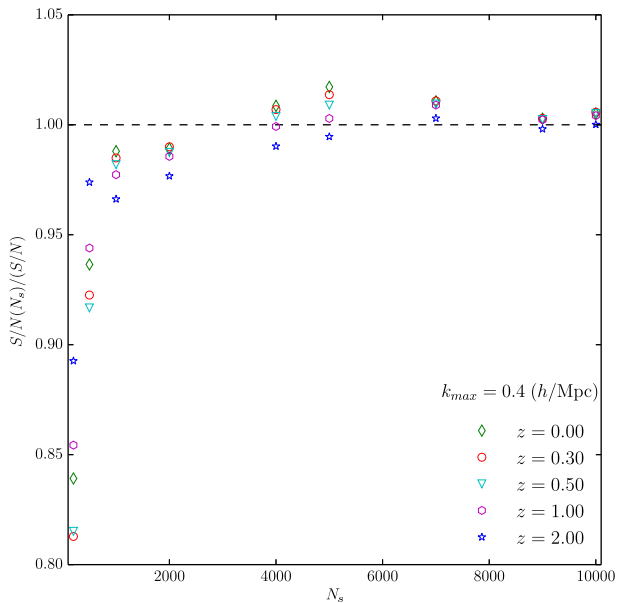}
\caption{Signal-to-noise of power spectrum measurements at $k_{\rm max}=0.40 \,h\,\textrm{Mpc}^{-1}$ estimated from sub-samples of Set A with mass resolution correction and normalised to the signal-to-noise of the full Set A for $z=0,0.3,0.5,1$ and $2$ respectively. Convergence at the per cent level is achieved for $N_s>4000$.}
\label{signoise_conv}
\end{center}
\end{figure}

In Fig.~\ref{signoise_conv} we plot the signal-to-noise at $k_{\rm max}=0.40 \,h\,\textrm{Mpc}^{-1}$ (a scale on the plateau of the S/N) as function of the number of realisations at different redshifts. As we can see the signal-to-noise converges for $N_{\rm s}>4000$ at all the redshifts within a few per cent.

Here, it is worth noticing that the signal-to-noise has been estimated using an unbiased estimator of the precision matrix, Eq.~(\ref{unbiasedinvcov}). However, this is only valid in the Gaussian case for which the probability distribution of the inverse covariance is given by the inverse-Wishart distribution. As we have shown in the previous Section, at low redshift and high $k$ the statistics of the matter power spectrum deviates from that of Gaussian random density field. Hence, it is reasonable to expect a departure of the precision matrix from the inverse-Wishart distribution at these scales and redshifts. This is particularly relevant for cosmological model parameter estimations and studies along the line of the analysis by \citep{taylor13,taylor14} using the DEUS-PUR simulations can be particularly informative. A study which we leave to future work.

\section{Conclusion}\label{conclu}
The covariance matrix of the matter power spectrum is an essential ingredient to infer unbiased cosmological parameter constraints from measurements of the clustering of matter in the universe. The next generation of galaxy surveys will precisely measure the matter power spectrum across a wide range of scales. This demands for an accurate estimation of the covariance matrix. Here, we have tackled this task using a large ensemble of numerical N-body simulations from the DEUS-PUR project. First, we have assessed the impact of numerical systematic uncertainties due to mass resolution effects using a reduced sample of higher resolution simulations. We have provided an empirical statistical method to correct for this source of non-Gaussian error on the covariance matrix. By taking advantage of the large statistics of the DEUS-PUR simulations we have finely sampled the power spectrum probability distribution. We have found a non-vanishing skewness which deviates from the expectations of a Gaussian density field at low redshifts ($z\le 0.5$) with the amplitude of the deviations increasing on scales $k\gtrsim 0.25 \,h\,\textrm{Mpc}^{-1}$. The non-Gaussian errors resulting from the non-linear regime of gravitational collapse mildly affect the spectrum on BAO scale, in contrast they become important at smaller scales. This suggests that an unbiased estimate of the ensemble averaged band power from finite volume surveys at these scales and redshifts may require the full probability distribution of the matter power spectrum. The ensemble of N-body simulations from the DEUS-PUR project can provide a valuable support to these future analyses.

\section*{Acknowledgements}
We would like to thank Tom Kitching for useful discussions. This work was granted access to HPC resources of IDRIS through allocations made by GENCI (Grand \'Equipement National de Calcul Intensif) in the framework of the `Grand Challenge' DEUS-PUR on the machine ADA. We acknowledge support from the DIM ACAV of the Region \^Ile-de-France. The research leading to these results has received funding from the European Research Council under the European Union's Seventh Framework Programme (FP/2007-2013) / ERC Grant Agreement n. 279954.

\appendix
\section{Mass resolution correction to matter power spectrum} \label{appendix}
Here, we derive the mass resolution correction to the matter power spectra estimated from Set A. 

We assume that the corrected power spectrum estimator, $\hat{P}^{\rm corr}_{\textrm{A}}$, and that of the lower resolution simulations, $\hat{P}_{\textrm{A}}$, are related by a simple linear transformation: 
\begin{equation}\label{pcorrtransf}
\hat{P}^{\rm corr}_{\textrm{A}}=a\, \hat{P}_{\textrm{A}} + b.
\end{equation}
The goal here is to find a correction that maps each of the $\hat{P}_{\textrm{A}}$ from the pdf of Set A, $f(\hat{P}_{\textrm{A}})$, into the one of Set B, $f(\hat{P}_{\textrm{B}})$. Since the proposed correction has two parameters, we only need the first two moments of $f(\hat{P}_{\textrm{B}})$ to correct the spectra of Set A. In principle one can assume higher order corrections, but then higher moments of $f(\hat{P}_{\textrm{B}})$ are needed for the computation, and the statistics of our sample is not sufficient to resolve them.\\
 We determine the coefficients $a$ and $b$ by imposing that the average $\bar{P}^{\rm corr}_{\textrm{A}}=\bar{P}_{\textrm{B}}$ and the variance $\sigma^2_{\hat{P}^{\rm corr}_{\textrm{A}}}=\sigma^2_{\hat{P}_{\textrm{B}}}$. Those conditions translate in the system:
\begin{equation}
\bar{P}_{\textrm{B}}=a\,\bar{P}_{\textrm{A}}+b,
\end{equation}
\begin{equation}
\sigma^2_{\hat{P}_{\textrm{B}}}=a^2\, \sigma^2_{\hat{P}_{\textrm{A}}},
\end{equation}
from which we finally obtain Eq. (\ref{pcorr}):
\begin{equation}
\hat{P}^{\rm corr}_{\textrm{A}}= \frac{\sigma_{\hat{P}_{\textrm{B}}}}{\sigma_{\hat{P}_{\textrm{A}}}} \hat{P}_{\textrm{A}}+\bar{P}_{\textrm{B}}-\frac{\sigma_{\hat{P}_{\textrm{B}}}}{\sigma_{\hat{P}_{\textrm{A}}}}\bar{P}_{\textrm{A}}.
\end{equation} 

The standard deviation of the power spectra from Set B is very noisy, and so is the ratio $\sigma_{\hat{P}_{\textrm{B}}}/\sigma_{\hat{P}_{\textrm{A}}}$. We find convenient to smooth out this noise and assume a fourth grade polynomial fitting function of $k$, such that $\sigma_{\hat{P}_{\textrm{B}}}/\sigma_{\hat{P}_{\textrm{A}}}\equiv\alpha\,k^4+\beta\,k^3+\gamma\,k^2+\delta\,k + 1$ where $\alpha,\beta,\gamma,\delta$ are the fitting parameters that we obtain by fitting the polynomial function to the numerical ratio of the standard deviations. We impose that on the large linear scales, e.g. at $k_{\rm min}$ this ratio tends to unity since in this regime there is no difference between Set A and B. 

\begin{figure}
\begin{center}
\includegraphics[scale=0.5]{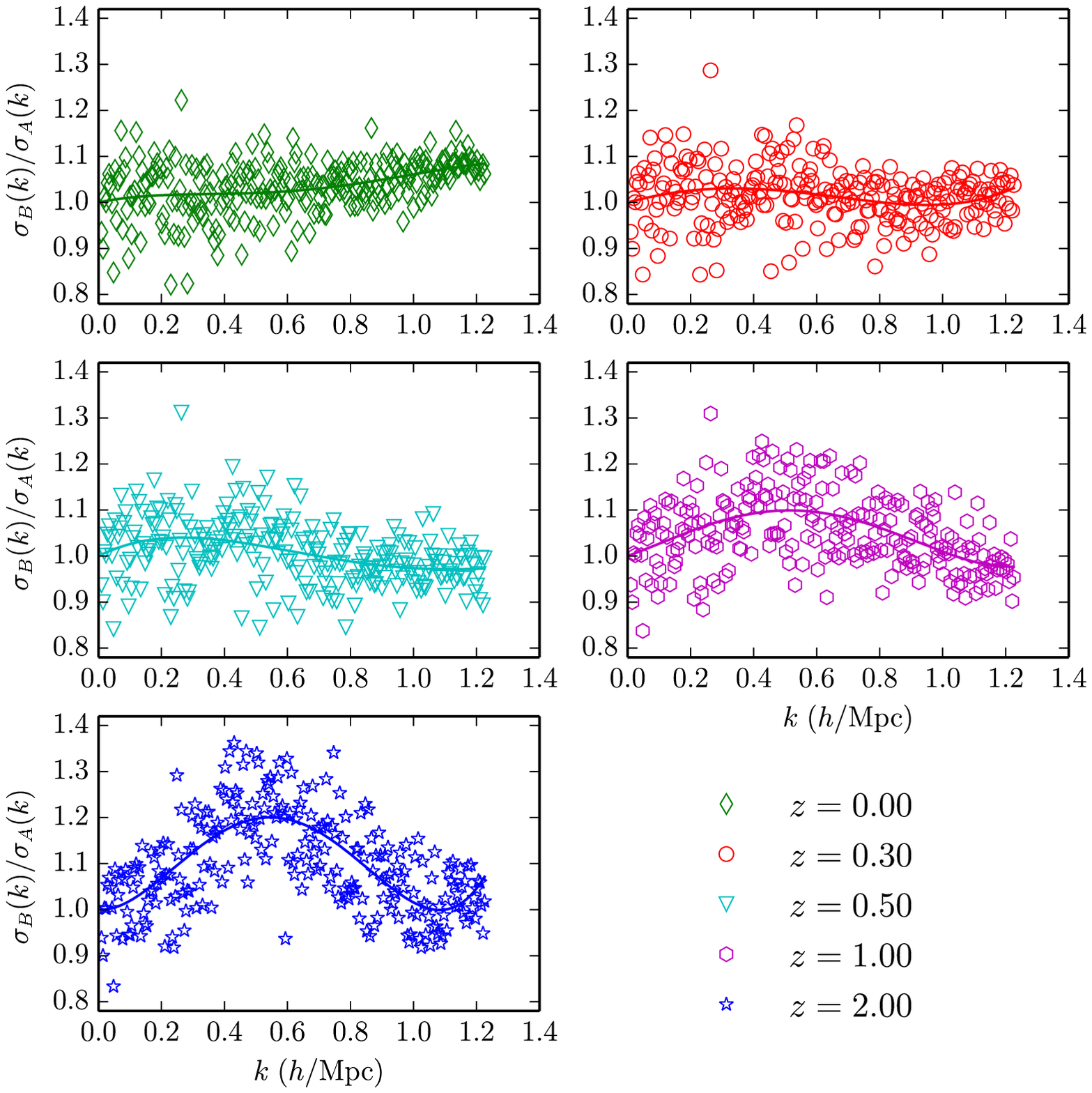}
\caption{Ratio of the standard deviation of the spectra from Set B and A, $\sigma_{\hat{P}_{\textrm{B}}}/\sigma_{\hat{P}_{\textrm{A}}}$ as function of $k$ at $z=0,0.3,0.5,1$ and $2$ respectively. The solid lines are the best-fitting smoothing functions.}
\label{smooth}
\end{center}
\end{figure}

In \fig~\ref{smooth} we plot $\sigma_{\hat{P}_{\textrm{B}}}/\sigma_{\hat{P}_{\textrm{A}}}$ and the best-fitting smoothing function at $z=0,0.3,0.5,1$ and $2$ respectively. The best-fitting values of the parameters are quoted in Table~\ref{tab:bestfit}.

\begin{table}
\centering
\begin{tabular}{c|c|c|c|c}
\ $z$ & $\alpha$ & $\beta$ & $\gamma$ & $\delta$\\
\hline
\ $0$ & $-0.156$ & $0.569$ & $-0.442$ & $0.146$\\
\ $0.3$ & $0.098$ & $-0.039$ & $-0.290$ & $0.188$\\
\ $0.5$ & $-0.410$ & $0.695$ & $-0.895$ & $0.343$\\
\ $1$ & $0.266$ & $1.026$ & $0.280$ & $0.252$\\
\ $2$ & $2.240$ & $-5.272$ & $2.924$ & $-0.051$ \\
\end{tabular}
\caption{Best-fitting values for the parameters $\alpha,\beta,\gamma$ and $\delta$.}
\label{tab:bestfit}
\end{table}

\end{document}